\documentclass[conference]{IEEEtran} 
\IEEEoverridecommandlockouts
\newcommand{\scheme}{HERL\xspace}

\def\BibTeX{{\rm B\kern-.05em{\sc i\kern-.025em b}\kern-.08em
    T\kern-.1667em\lower.7ex\hbox{E}\kern-.125emX}}
\usepackage[utf8]{inputenc} % allow utf-8 input
\usepackage{amsmath}
\usepackage{xspace}
\usepackage{bbm}
\usepackage{enumitem}
\usepackage{tcolorbox}
\usepackage{graphicx} % Required for inserting images
\usepackage{algorithm}
\usepackage[noend]{algpseudocode}
\usepackage{graphicx}
\usepackage[T1]{fontenc}    % use 8-bit T1 fonts
\usepackage{amsmath} 
\usepackage[hidelinks]{hyperref}
\usepackage{url}            % simple URL typesetting
\usepackage{booktabs}       % professional-quality tables
\usepackage{amsfonts}       % blackboard math symbols
\usepackage{nicefrac}       % compact symbols for 1/2, etc.
\usepackage{microtype}      % microtypography
\usepackage{subfigure}
\usepackage{multicol,multirow}

\title{HERL: Tiered Federated Learning with Adaptive Homomorphic Encryption using Reinforcement Learning}

\author{
    Jiaxang Tang\textsuperscript{1,*}, Zeshan Fayyaz\textsuperscript{2,*}, Mohammad A. Salahuddin\textsuperscript{2}, Raouf Boutaba\textsuperscript{2}, Zhi-Li Zhang\textsuperscript{1}, and Ali Anwar\textsuperscript{1} \\
    \textsuperscript{1}Computer Science and Engineering, University of Minnesota-Twin Cities, Minneapolis, MN, USA \\
    \textsuperscript{2}David R. Cheriton School of Computer Science, University of Waterloo, Ontario, Canada \\
    \textsuperscript{*}These authors contributed equally to this work. \\
    Email: \{tang0836, zhang089, aanwar\}@umn.edu, \{z3fayyaz, mohammad.salahuddin, rboutaba\}@uwaterloo.ca
}

\begin{document}

\maketitle

\pagestyle{plain} 

\begin{abstract}
Federated Learning is a well-researched approach for collaboratively training machine learning models across decentralized data while preserving privacy. However, integrating Homomorphic Encryption to ensure data confidentiality introduces significant computational and communication overheads, particularly in heterogeneous environments where clients have varying computational capacities and security needs. In this paper, we propose \scheme, a Reinforcement Learning-based approach that uses Q-Learning to dynamically optimize encryption parameters, specifically the polynomial modulus degree, $N$, and the coefficient modulus, $q$, across different client tiers. Our proposed method involves first profiling and tiering clients according to the chosen clustering approach, followed by dynamically selecting the most suitable encryption parameters using an RL-agent. Experimental results demonstrate that our approach significantly reduces the computational overhead while maintaining utility and a high level of security. Empirical results show that \scheme improves utility by 17\%, reduces the convergence time by up to 24\%, and increases convergence efficiency by up to 30\%, with minimal security loss.
\end{abstract}


\section{Introduction}
Federated Learning (FL) has emerged as a collaborative machine learning (ML) technique that enables multiple clients to train a shared model while keeping their individual data private and secure \cite{mcmahan2017communication}. Since its initial development by Google in 2018, substantial research has been performed in optimizing various areas of the FL pipeline \cite{kairouz2019advances}. Overhead reduction techniques have been deployed in the domain of client selection and data selection \cite{li2020federated}. 
However, the need for more sophisticated security measures, coupled with the massive amount of sensitive data IoT devices are generating every day adds substantial complexity and overhead \cite{xu2019federated}. 

\begin{figure}
    \centering
    \subfigure[Parameter plan on latency.]{\includegraphics[width=0.48\linewidth]{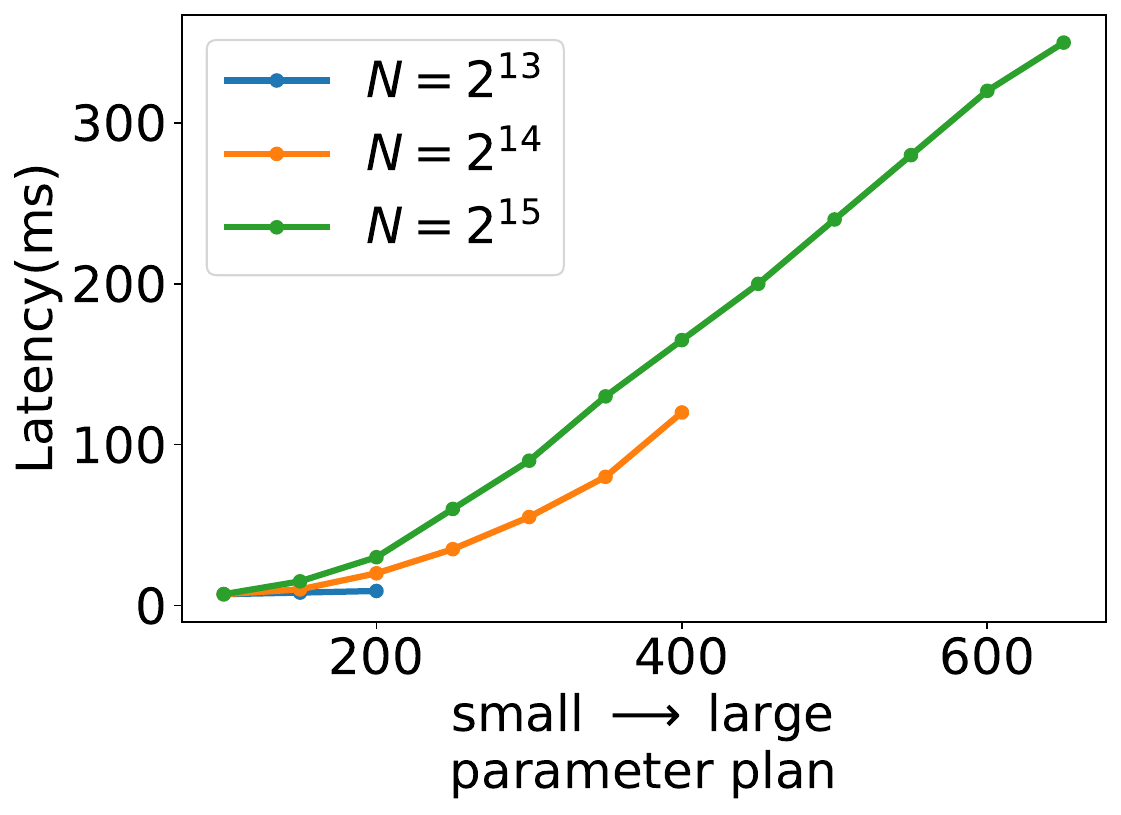}}
    \subfigure[Parameter plan on security.]{\includegraphics[width=.48\linewidth]{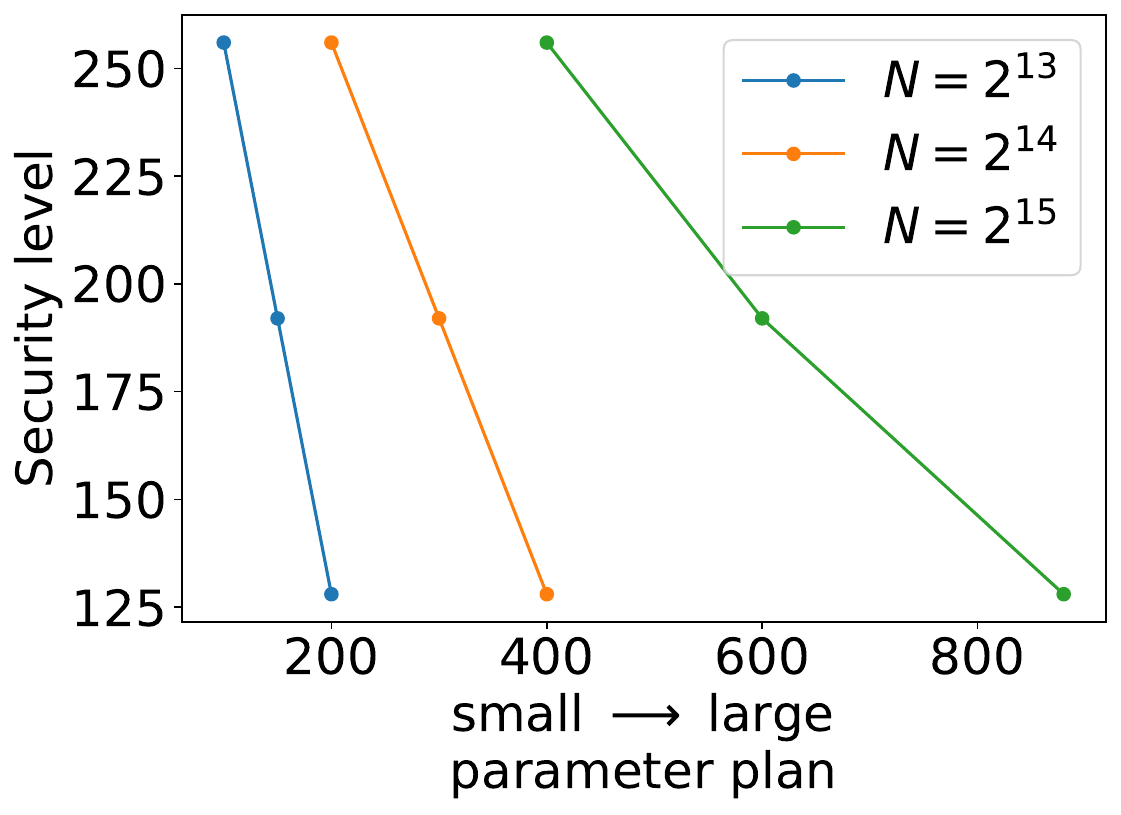}}
    \vspace{-.2cm}
    \caption{The influence HE parameter plan on latency and security}
    \label{fig:paramplan}
    \vspace{-.5cm}
\end{figure}

Many privacy-preserving FL techniques have been extensively studied, including secure aggregation \cite{bonawitz2016practicalsecureaggregationfederated}, differential privacy \cite{wei2019federatedlearningdifferentialprivacy}, and homomorphic encryption (HE) \cite{HE, LI2020106854}. HE allows computation on the ciphertexts and requires fewer communication interactions which is particularly suitable for cross-device settings. 
HE brings forth many advantages, such as allowing updates to remain encrypted throughout the FL process, and ensuring that sensitive data never gets exposed, even to the central server \cite{gentry2009fully}. However, HE in FL introduces significant computational overhead and inherent error. This overhead can be a substantial barrier in real-world settings, given the heterogeneity of the capacity of client devices, such as computational power, resources, and varying security requirements \cite{mo2021efficient}. Furthermore, the complicated correlations of HE parameters have impeded the relevant research study to overcome this heterogeneous setting. For each HE application, there is a parameter plan, consisting of polynomial modulus $N$ and coefficient modulus $q$, to setup the encryption scheme. The general influence of a parameter plan on latency and security is shown in Figure \ref{fig:paramplan}. In general, an increasing parameter plan of HE leads to larger latency and a higher security level. The complexity increases due to the limitation imposed by $N$, which restricts the upper bound of the parameter plan scheme that can be supported. Note that HE introduces quantization error during the encoding step required to scale float numbers to finite fields. The values of $q$ linearly restrict the maximum scale level. Correspondingly, the influence of an HE parameter plan on precision is simple, as it is linearly scaled.

Current approaches for FL with HE often adopt a uniform strategy across all participating clients, which is inefficient given the diverse computational capacities and security requirements of clients \cite{mo2021efficient}. For instance, some clients (e.g., government users) manage highly sensitive data, therefore needing robust encryption, while others (e.g., general residents) may prioritize computational efficiency with moderate security needs \cite{tan2022federated}. Prior research does not take the heterogeneity and diversity of security needs of clients into consideration \cite{zhang2024hybrid}. 
As shown in the previous results, different parameter plans will have distinct influences on the FL system. There exists a clear gap between the uniform HE parameter plan and varying client characteristics, which is inefficient and impractical for large-scale FL settings \cite{li2021privacy}. The goal of our work is to make ease of HE parameter selection to adapt to the heterogeneous environment while balancing security and performance.

In this paper, we present a novel reinforcement learning (RL)-based agent, named \scheme, designed to address the outlined concerns by intelligently selecting the HE parameter plan. This approach enables an effective trade-off between latency, utility, and security in FL, optimizing performance across diverse client conditions. Specifically, we begin with profiling the clients based on their resources (including CPU, GPU, network bandwidth, etc) and security requirements, and tiering according to chosen clustering approach. 
By populating a Q-table with $q$ and $N$ values using exploration-exploitation strategy, we then use RL to intelligently assign ($q$, $N$) values to each sub-cluster. We design a new reward function to balance latency, utility, and security. Based on the comprehensive evaluation, \scheme could enhance the efficiency of FL training by up to 24\% while meeting the requirements of utility and security.

The main contributions are as follows: 

\begin{itemize} 
    \item We propose a novel RL-agent based approach that dynamically selects optimal HE parameters, generic to any FL clustering approach, adapting to client-specific requirements to balance security and performance in FL. 
    \item A better trade-off between latency, utility, and security by leveraging an adaptive approach, ensuring that computational overhead is minimized while maintaining the necessary level of data protection in FL. 
    \item Comprehensive experiments are conducted to evaluate the performance of \scheme. \scheme is shown to achieve significant improvement on training efficiency, by up to 24\%.
\end{itemize}

Furthermore, we support those contributions by answering the following key research questions throughout the paper:

\begin{tcolorbox}[colback=white!95!gray, colframe=gray!50!black, title=Research Questions (RQs), fonttitle=\bfseries, coltitle=black, colbacktitle=white!50!gray, boxrule=0.5mm, rounded corners=all]
\begin{enumerate}[label=\arabic*., leftmargin=*]
\item\label{ite:fiv} \textbf{Influence: }How will the parameters of HE influence FL performance? And, how could we correctly use HE for FL applications?
\item\label{ite:fir} \textbf{Generalize: }How do we generalize the clustering step for heterogeneous settings?
\item\label{ite:sec} \textbf{Optimize: }How can we optimize the trade-off among utility, computational overhead, and security requirements in FL with HE? \label{rq2} %reward we balance the trade-off between round efficiency with security sacrifice
\item\label{ite:thi} \textbf{Trade-off: }How can RL be effectively applied to dynamically adjust HE parameters across different client tiers? %We train Q-learning agent based on the training process, refer to Float; or leave for me.
\item\label{ite:fou} \textbf{Effectiveness: }How does the RL-based approach perform, and do the selected parameters ensure a better trade-off?

% \item\label{ite:fiv} Will the pre-trained RL agent generalize to other cluster based FL approaches? Otherwise, how do we fine-tune?%experiment if get time
\end{enumerate}
\end{tcolorbox}

\section{Preliminary}
In this section, we provide a comprehensive overview of the foundational concepts and mechanisms behind FL and HE, highlighting the integration of HE within FL systems and its implications for privacy-preserving ML.
\subsection{Federated Learning}
FL is a distributed ML paradigm wherein a central server orchestrates the training of a global model by aggregating updates from multiple decentralized clients \cite{mcmahan2017communication}.
The FL process generally unfolds in three phases. Initially, the central server selects a subset of clients to participate in each training round. This selection is crucial as the efficiency and accuracy of the training process can be significantly influenced by the availability of clients, their computational power, and network conditions \cite{kairouz2019advances}. Following this, the selected clients train the global model locally on their respective datasets. The goal during this phase is to minimize a local loss function \( \mathcal{L}_i(w; \mathcal{D}_i) \) concerning the model parameters \( w \), where \( \mathcal{D}_i \) represents the local dataset of client \( i \). After local training, the resulting model parameters \( w_i \) are sent back to the server \cite{li2020federated}.
The final phase involves the server aggregating the updates from all participating clients to refine the global model. Typically, this is done using Federated Averaging (FedAvg), where the server computes a weighted average of the model parameters \cite{mcmahan2017communication}:

\begin{equation}
w \leftarrow \sum_{i=1}^{N} \frac{|\mathcal{D}_i|}{|\mathcal{D}|} w_i
\end{equation}
where \( |\mathcal{D}_i| \) represents the size of the local dataset of client \( i \), and \( |\mathcal{D}| = \sum_{i=1}^{N} |\mathcal{D}_i| \) is the total size of all datasets involved in that round. This aggregation ensures the global model benefits from the diverse datasets while preserving data privacy \cite{konevcny2016federated}.

However, the variability in data and resource availability across clients can negatively impact model performance and result in inconsistent response times in cross-device FL, leading to significant bottlenecks, such as the straggler problem. Factors such as low computational power, bandwidth constraints, and battery limitations contribute to this issue \cite{bonawitz2019towards}. Without adaptive optimizations that account for these differences, stragglers can hinder system efficiency, extend convergence time, and lead to resource dropouts, where clients with limited resources are unable to complete training tasks \cite{li2020federated}.

\subsection{Homomorphic Encryption}
HE is a cryptographic technique that allows computations to be performed directly on encrypted data, producing an encrypted result that, when decrypted, matches the outcome of operations performed on the plaintext \cite{gentry2009fully}. This technique is critical in privacy-preserving ML, particularly in FL, where model updates are encrypted before being sent to the server, ensuring data privacy throughout the training process \cite{aono2017privacy}.

Among the various HE schemes, CKKS \cite{ckks} stands out as a leveled HE scheme designed for approximate arithmetic, balancing computational efficiency and security \cite{cheon2018bootstrapping}. Other HE schemes, such as BFV \cite{bfv} and BGV \cite{bgv}, offer exact arithmetic but at the cost of higher computational demands \cite{brakerski2011fully}. CKKS is particularly well-suited for FL scenarios where privacy must be preserved without sacrificing computational efficiency \cite{cheon2017homomorphic}.  CKKS allows efficient parallel operations on encrypted data, making it well-suited for applications like ML where ciphertext size is extremely large \cite{cheon2017homomorphic}. Note that we implicitly use HE to represent the CKKS scheme throughout this paper.

The CKKS process consists of setup, key generation, encryption, computation, and decryption.
In the setup phase, parameters such as the polynomial modulus degree $N$, coefficient modulus $q$, and scaling factor $\Delta$ are chosen. The polynomial modulus degree defines the size of ciphertexts and computational complexity, while the coefficient modulus determines the precision of computations. The scaling factor is applied to plaintexts to quantize the float point numbers to the encoding field. The key generation process involves creating several keys: the secret key ($sk$) for decryption, the public key ($pk$) for encryption, a relinearization key to manage ciphertext size after multiplication, and a rotation key for homomorphic rotations of encrypted vectors.
In the encryption phase, plaintext values are encoded into vectors of complex numbers. After encoding, the plaintext is scaled by the scaling factor, ensuring precision in future operations. The public key is then used to encrypt the encoded plaintext, generating a ciphertext ready for computation.

\begin{equation}
    E(m_1) \odot E(m_2) = E(m_1 \odot m_2)
\end{equation}

where $E$ is the encryption function and $\odot$ denotes operations, such as addition or multiplication on the ciphertext field for HE \cite{gentry2009fully}.
With the secret key, clients can decrypt the ciphertext, recovering the encoded result. The final decoding step removes the scaling factor and outputs an approximate result of the original plaintext.

\begin{figure}
    \centering
    \subfigure[Impact of $q$ on latency.]{\includegraphics[width=0.48\linewidth]{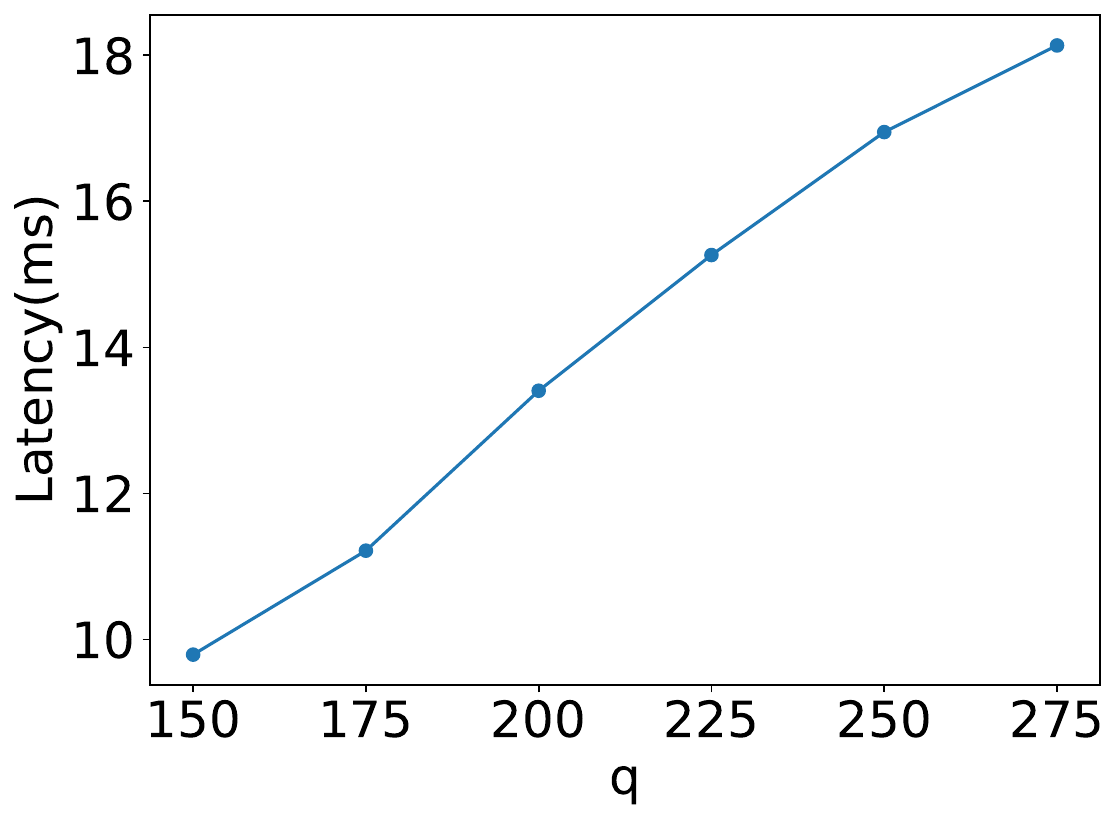}}
    \subfigure[Impact of $q$ on utility.]{\includegraphics[width=0.48\linewidth]{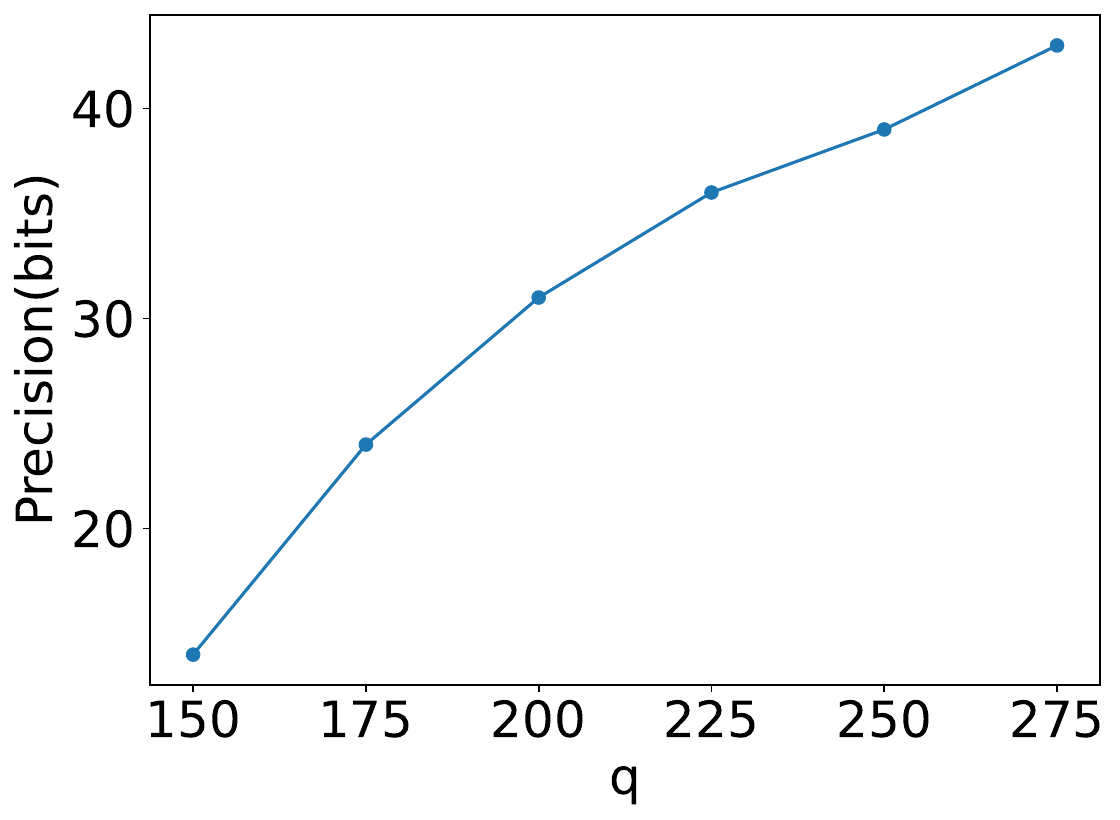}}\\
    \subfigure[Impact of $N$ on latency.]{\includegraphics[width=0.48\linewidth]{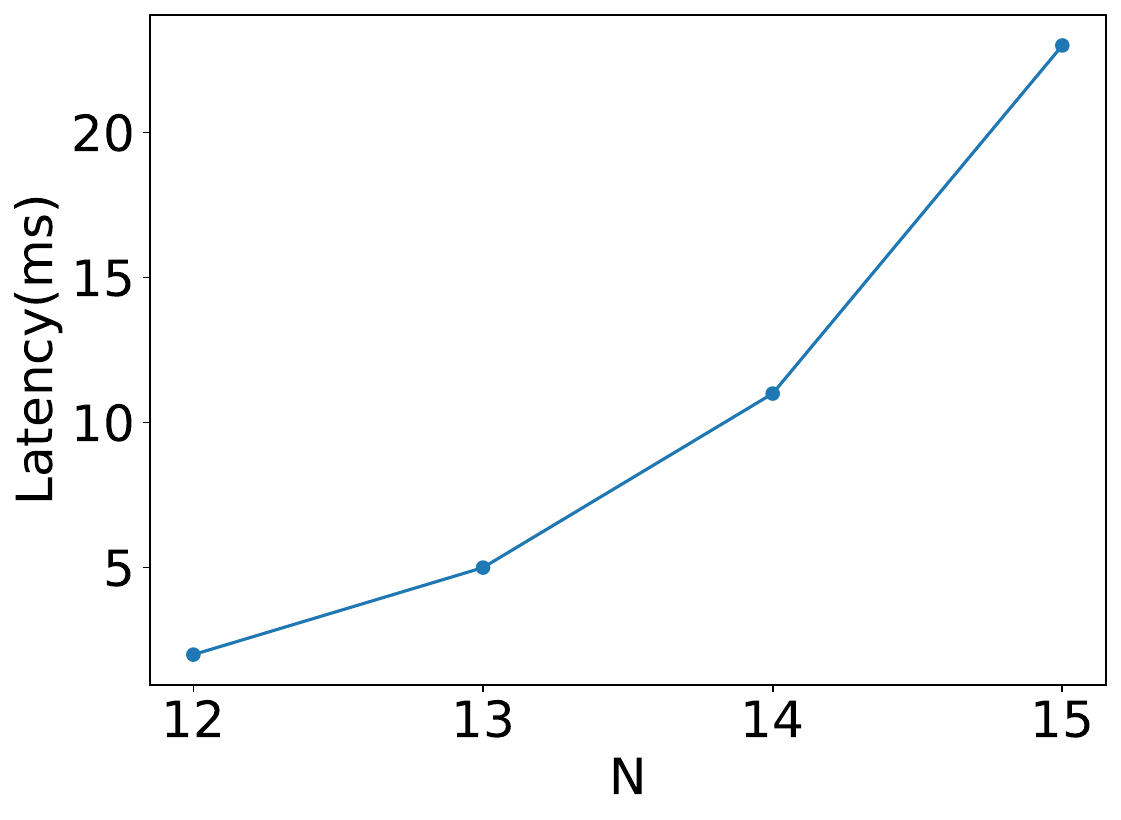}}
    \subfigure[Impact of $N$ on utility.]{\includegraphics[width=0.48\linewidth]{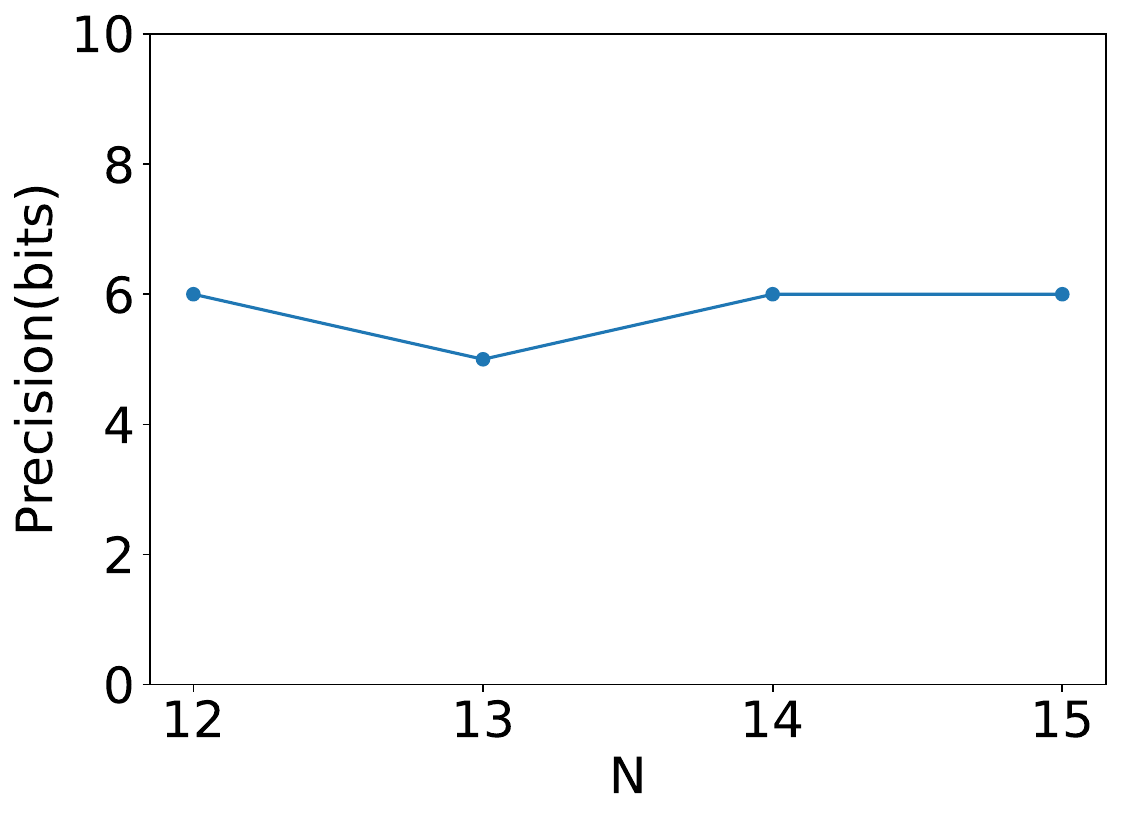}}\\
    \caption{Impact of different HE parameter plans on latency and precision.}
    \label{fig:HE}
\end{figure}

\noindent\textbf{HE parameter plan.} The parameter choice for the pair ($q,N$) is named the parameter plan for HE. This pair correlates to all three aspects of the HE application, including utility (ML accuracy), latency, and security.
The influences of $q$ and $N$ are presented in Figure \ref{fig:HE}. Specifically, as shown in Figure \ref{fig:HE} (a) and (b), parameter $q$ is proportional to the precision and latency, that is, larger $q$ values have better precision but at the cost of larger latency. In Figure \ref{fig:HE} (c) and (d), the parameter $N$ is exponential to the latency, but doesn't directly influence the precision. However, $N$ correlates with $q$ by limiting the largest bound of $q$. That is, a larger $N$ would allow a larger $q$, hence affording better precision. Note, that the utility of HE depends on the HE application (e.g., here, utility represents the model accuracy when we apply HE to FL/ML).

\noindent\textbf{Threat model.} We focus on a semi-honest (honest-but-curious) threat model, all parties involved in the HE protocol are assumed to follow the protocol correctly, but they may attempt to learn additional information by inspecting the ciphertexts or intermediate results. This model ensures confidentiality against adversaries who respect the protocol but may try to infer sensitive data from the encrypted messages. However, we do not consider the malicious model, especially when clients proactively send the poisoning models (poison attack) to attack the global model performance.

\subsection{Federated Learning with Homomorphic Encryption}

Combining the functionality of FL with HE offers a promising approach for privacy-preserving collaborative model training by allowing participants to encrypt their locally updated models before sharing them with a central server \cite{aono2017privacy}. The server can then aggregate these encrypted models without decrypting them, preserving the privacy of each participant's data \cite{gentry2009fully}. Integrating FL with HE includes the following steps. First, we begin with distributing the keys - that is, in FL with HE, the distribution of cryptographic keys involves distributing either the entire or parts of the private key to the clients, while the public key is shared with all participants prior to the commencement of training \cite{cheon2017homomorphic}. Next, we commence local training where clients get the current global model, then locally train the global model with their own data. Following this, clients use HE to encrypt their updated models. Finally, aggregation occurs, where clients send their encrypted model to the server, which averages all received local encrypted models \cite{aono2017privacy}. 
This iterative process continues until the model reaches convergence.

Despite its advantages, FL with HE faces significant challenges. For example, stragglers - clients with lower computational power or network bandwidth can slow down the training process, leading to inefficiencies and delays \cite{bonawitz2019towards}. Furthermore, the non-IID (Independent and Identically Distributed) nature of data across clients can exacerbate disparities in model performance, thereby reducing overall model utility and fairness \cite{zhu2021federated}. The use of HE, while enhancing privacy, introduces additional computational overhead and potential errors, which further complicates the design of FL systems \cite{mo2021efficient}. These challenges highlight the need for advanced techniques to address the complexities arising from heterogeneous resources and varying security requirements among clients in FL environments.

\section{Related work}
In this section, we briefly review the related works on FL, addressing privacy and heterogeneity problems.

\textbf{Clustered Federated Learning.} Client heterogeneity, whether in data distribution or security requirements, presents a significant challenge in optimizing FL systems \cite{kairouz2019advances}, which can hinder model performance \cite{zhu2021federated}. Clustered FL approaches have been explored to tackle these challenges, IFCA \cite{ifca} utilizes a random clustering approach to overcome the challenges that arise from data heterogeneity. FL+HC \cite{briggs2020federated} introduces Hierarchical Clustering, which clusters clients based on the similarity of their local updates to improve utility and reduces the number of communication rounds required for convergence. \cite{briggs2020federated}. TiFL \cite{chai2020TiFL} utilizes tiered-based clustering and selects clients based on utility to significantly improve performance under heterogeneous settings. More recent work, such as Auxo \cite{liu2023auxo} proposes to use gradient distribution to cluster clients to improve the convergence efficiency. However, these clustering methods are not based on privacy-preserving FL, thereby disregarding privacy concerns.

\textbf{Privacy-Preserving Federated Learning}. Existing decentralized privacy-preserving techniques have been widely explored within FL, including MPC \cite{bonawitz2016practicalsecureaggregationfederated,fereidooni2021safelearn,so2022lightsecagg}, differential privacy \cite{truex2020ldp,triastcyn2019federated}, HE \cite{hosseini2021secure}, and blockchain \cite{GAIN}. Based on the unique characteristics of HE and the lack of research on HE within FL frameworks, we focus on the FL+HE system. The work secureMF \cite{chai2020secure} applies HE to matrix factorization under a federated setting to guarantee the security of aggregation. Furthermore, BatchCrypt \cite{zhang2020batchcrypt} proposes a more sophisticated quantization scheme for HE to improve efficiency when applying HE to FL. xMK-CKKS \cite{ma2022privacy} designed multi-key HE to improve the security of the secret key, especially under the FL's multi-party setting. A hybrid privacy-preserving FL approach \cite{truex2019hybrid} integrates differential privacy (DP) with threshold-HE to overcome the inference vulnerability of HE and utility loss of DP. More recent work, such as adaptive batch HE framework for cross-device FL \cite{adapthe} explores HE under heterogeneous FL settings which consider a simple additional HE scheme (Paillier \cite{paillier1999public}) and adjusts the key size for clients with different resources. However, this scheme does not support multiplication and is not efficient compared to the more advanced HE scheme targeting the current big data setting.

In comparison, our work focuses on optimizing privacy-preserving FL systems that incorporate an advanced HE scheme (CKKS \cite{ckks}). While previous research has often overlooked the varying security requirements across clients, we aim to fill this gap. Similar to clustered FL approaches, we tackle the challenges posed by client heterogeneity in terms of efficiency and performance. However, our approach uniquely focuses on reducing the computational overhead introduced by HE, which is becoming increasingly problematic in complex FL systems. We utilize an RL-agent (specifically Q-learning) to dynamically adjust HE parameters across client tiers, balancing security and computational efficiency in a way that adapts to the heterogeneous nature of client devices and their varying security requirements.

\section{Motivation}
Integrating HE into FL is crucial for maintaining data confidentiality but introduces significant computational and communication overheads, especially in real-world heterogeneous environments \cite{mo2021efficient}. To address this challenge, we propose a tier-based approach that enables adaptive HE, optimizing parameter settings to match the characteristics of individual clients. This approach helps mitigate the overheads by tailoring encryption levels based on client capabilities.

\subsection{Influence of heterogeneity and HE parameter plan.} 
We begin by introducing the influence of client-side heterogeneity (including resources and security) and the HE-side distinct parameter plans through a case study.

\begin{figure}[!ht]
\hspace*{-0.5cm}
\includegraphics[width = .5\textwidth]{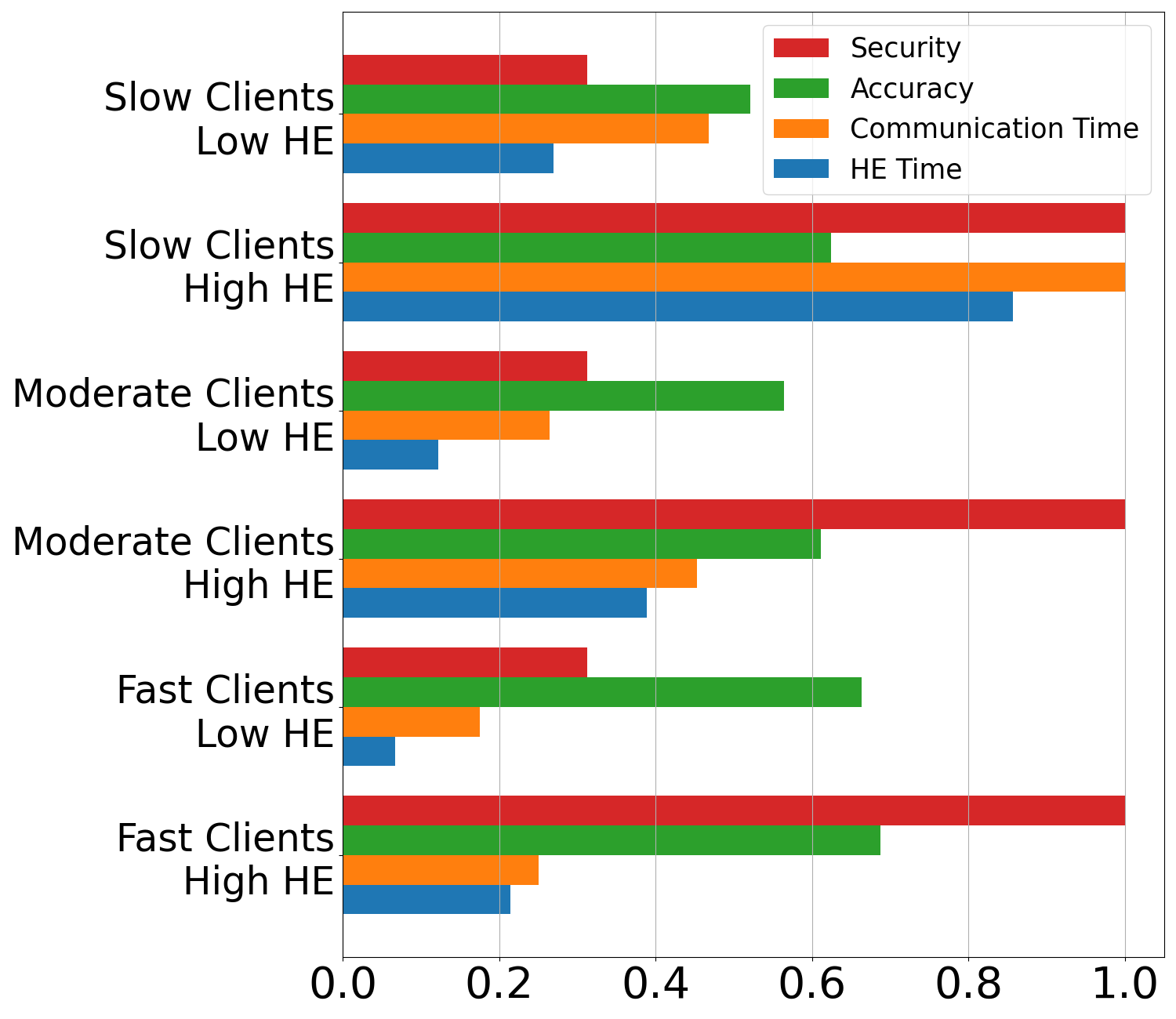}
\caption{Comparing the normalized Security (red) and Accuracy (green), Communication time (orange), and HE time (blue) across six configurations, categorized by client speed (Fast, Moderate, Slow) and HE parameters (High, Low).}
\label{fig:highHEvslowHE}
\end{figure}

\textit{Setup.} We conducted an initial exploration by running FL on the CIFAR-10 dataset \cite{CIFAR10} with 20 clients, each possessing different resource capacities by using real-world traces collected from \cite{bonawitz2019towards}. The HE parameter plan, as illustrated in Figure 2, shows that the impact of the parameter plan on FL performance scales almost linearly. Consequently, we selected two representative parameter configurations: low (with $q=100$, $N=13$) and high (with $q=200$, $N=14$). To account for resource heterogeneity, we categorized the clients as slow, moderate, or fast, based on their computing power and network bandwidth. Figure \ref{fig:highHEvslowHE} presents the results, revealing the correlation between FL performance, client characteristics, and the HE parameter plan. 

In terms of heterogeneity impact, even under the same parameter plan, we find varying performances among clients. As Figure \ref{fig:highHEvslowHE} presents, slow clients with fewer resources have a much larger latency in both computation and communication. Thus, during training, the FL system will be slowed down by those stragglers, especially clients with fewer resources.
In terms of heterogeneous security requirements, different HE parameter plans will directly influence the security guarantee for all the clients, since HE requires a uniform plan for all participants. A better privacy guarantee requires a higher HE parameter plan, which increases latency and better utility.

To address \textbf{RQ1: Influence}, which examines the impact of HE on FL clients we identify two key aspects: exacerbated heterogeneity and the trade-off between security and efficiency. As depicted in Figure \ref{fig:highHEvslowHE}, clients with limited computational resources, such as slow clients, face significant increases in both HE computation and communication time, particularly when high HE parameters are required to ensure security and utility. The size of the ciphertext generated by HE directly influences communication time, with larger ciphertexts, necessitated by higher HE settings, taking longer to transmit. This delay is especially pronounced for clients with slower network connections \cite{aono2017privacy}, exacerbating the straggler problem. On the other hand, while higher HE parameter settings enhance security and utility, they do so at the expense of increased computational and communication overhead.

\subsection{Adaptive HE parameter plan.} 
Based on the above results, it is clear that selecting a uniform plan for all clients with heterogeneous resources is not practical. Varying the HE parameter plan according to the resources of clients could mitigate the straggler problem significantly. 
Since HE needs a uniform parameter plan to do the ciphertext computation, clustering clients with similar characteristics is a natural way to assign different HE parameter plans among clients. To motivate our solution, we develop a case study on a two-cluster scenario to show the viability of an adaptive HE plan.

We split the clients into two parts, slow and fast, according to their resources. We then utilize three different HE parameter selection strategies to optimize the FL process, i.e., \textit{baseline, heuristic} and \textit{adaptive}. Specifically, in the \textit{baseline} approach, all clients use the same general recommended larger HE parameter plan $(N=14, q=200)$. While this configuration ensures a high level of security and better utility, it slows down the overall training process due to the slower clients \cite{mo2021efficient}. Furthermore, the \textit{heuristic} approach adopts a slow clients first strategy and uses a smaller HE parameter plan $(N=13, q=100)$ for all clients to catch the round time requirements. To mitigate this problem, we seek the potential of adaptive HE considering clients' characteristics, i.e., stragglers could use smaller HE plans to reduce the overhead.
Lastly, the \textit{adaptive} strategy, on the other hand, assigns lower HE parameter plan $(N=13, q=100)$ to slower clients and higher parameter plan $(N=14, q=200)$ to faster clients. Here, we use the uniform 256-bit privacy guarantee for all clients under all three settings. 

\begin{figure}[h!]
    \centering
    \includegraphics[width=0.75\linewidth]{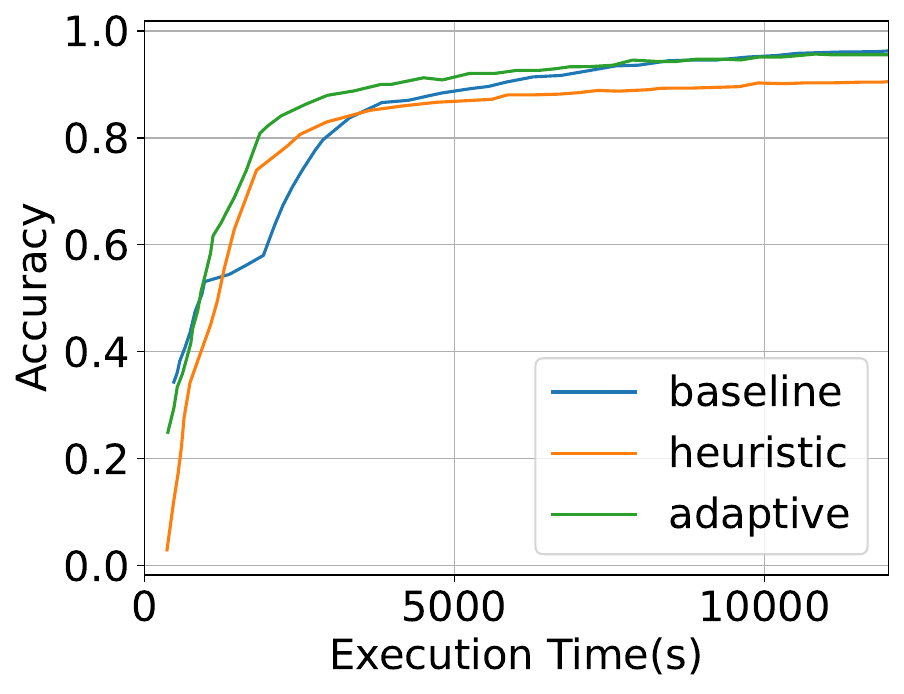}
    \caption{The relationship between test accuracy and total convergence time across three methods.} 
    \label{fig:adaptive}
\end{figure}

Figure \ref{fig:adaptive} presents the results of this comparison, showing that while the \textit{baseline} approach achieves high accuracy, it requires the longest convergence time due to the slower clients bottlenecking the process. The \textit{heuristic} approach significantly improves convergence speed by using smaller HE parameters for all clients, particularly in the early stages of training, but this comes at the cost of slightly lower accuracy. In contrast, the \textit{adaptive} approach strikes the best balance between accuracy and latency, achieving high accuracy in a shorter time compared to both \textit{baseline} and \textit{heuristic}.
This approach is the most effective as it speeds up the slower clients and limits the amount of error introduced by HE \cite{aono2017privacy}.

These experiments illustrate the critical need for balancing computational efficiency and utility in FL. More intensive HE parameters lead to a linear increase in training time, but better ML utility. This also underscores the importance of optimizing HE parameter assignment based on the specific characteristics of each client. The problem becomes more complicated when the security requirements are integrated. To overcome this difficulty, we utilize an RL-based approach to help optimize the HE parameter selection, thereby enabling better trade-offs among utility, latency, and security simultaneously.

\section{HERL}

\subsection{Overview}

\begin{figure}[t!]
    \centering
    \includegraphics[width=0.5\textwidth]{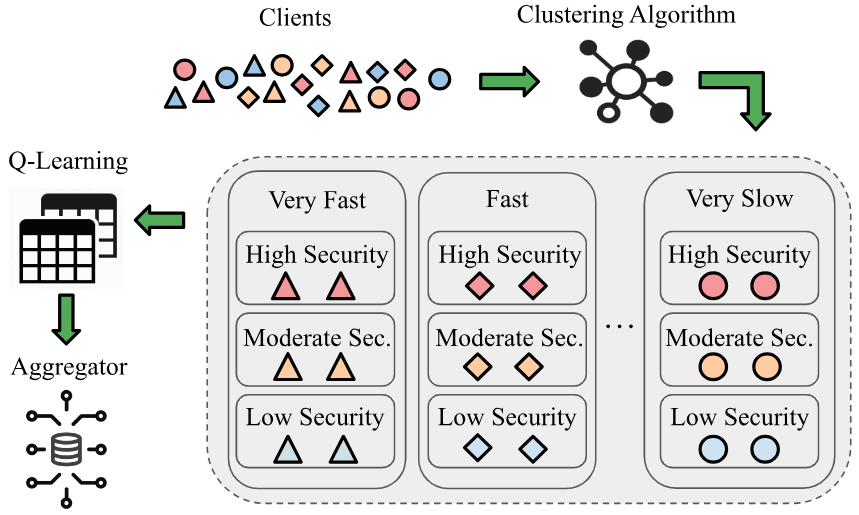}
    \caption{High level overview of \scheme.}
    \label{fig:architecture}
\end{figure}

An overview of the proposed \scheme architecture is given in Figure \ref{fig:architecture}. \scheme is designed to optimize FL systems by addressing the computational and communication overhead associated with HE. This architecture proposes a solution by dynamically adjusting HE parameters—specifically the coefficient modulus $q$ and the polynomial modulus degree $N$—to balance security and efficiency. The system is structured around three key components: profiling all clients to collect the characteristics of clients, clustering based on their capabilities, and an RL-agent (specifically Q-learning) to optimize HE parameters for each cluster. Throughout this section, we explain the details of \scheme whilst emphasizing the aforementioned research questions.

\subsection{Profiling and tiering}\label{sec:cluster}
We propose profiling and tiering to solve \textbf{RQ2: Generalize}. The architecture begins with a detailed profiling of all clients in the FL system. This profiling involves assessing each client’s computational resources, including CPU power, GPU availability, memory capacity, network bandwidth, utility distribution, and security requirements. Here, we aim to collect as many statistics as possible to make \scheme general enough to fit any existing clustering FL approach.

Based on this comprehensive profiling, clients are clustered into tiers based on the selected clustering approach. Here we utilize the hierarchical clustering FL+HC approach to illustrate the idea. To correlate with the three aspects of the FL/ML application mentioned previously, we tier clients based on their security requirements and latency. Specifically, we pre-define the number of criteria $\beta$ and tiers $K$. For each criterion, we split each existing sub-tier to $\sqrt[\leftroot{-2}\uproot{2}\beta]{K}$ further sub-tiers, until we iterate all of the criteria. For instance, we have $\beta=2$ criteria (security and latency) and $K=4$ intended total tiers. Firstly, we split clients according to the security requirements ([128,192)-bits, [192,256]-bits). Next, we split each of them further into two tiers, according to the latency of clients, i.e., we split clients within [128,192) to fast and slow tiers, we perform the same for the [192,256] tier.

Our scheme is designed to integrate seamlessly with existing clustered FL approaches, such as TiFL \cite{chai2020TiFL}, Auxo \cite{liu2023auxo}, IFCA \cite{ifca}, and others, by utilizing profiling information like client computational resources and network conditions to form clusters or sub-clusters. We propose hierarchical clustering as a flexible and efficient method for this tiering process, which organizes clients into layers based on similarities, enabling granular and scalable optimization. This approach allows for the merging of clients into clusters or the splitting of larger groups based on factors such as computational power, network bandwidth, or security requirements, resulting in better resource management and more precise parameter selection for each cluster. Additionally, our scheme is adaptable to various FL frameworks, making it easy to incorporate security needs into the clustering process when required. For instance, clients handling sensitive data can be further clustered based on security needs, ensuring appropriate HE settings are applied. This flexible integration makes our work easily extendable to future approaches.

\subsection{RL-based parameter selection}
Once all clients $\mathcal{C}$ are tiered to $K$ tiers $\{\mathcal{T}_t\}_{t=1}^K$, \scheme employs an RL mechanism to determine the optimal HE parameters for each sub-tier. The RL algorithm is responsible for selecting the parameter plan ($q$, $N$) values that will govern the level of encryption and the size of the ciphertext, respectively. The RL-agent explores different combinations of these parameters, learning over time which configurations offer the best trade-offs between security and performance for each client sub-tier. This dynamic adjustment is crucial for maintaining the efficiency of the FL process, as it ensures that each client is operating with HE settings that are tailored to its specific requirements, thereby minimizing unnecessary overhead while maintaining the desired level of data protection.

Specifically, for the RL-agent, we utilize a Q-learning approach, where the RL-agent builds a Q-table, with each entry linked to a specific state—reflecting the current conditions of a client sub-tier, such as the average of available resources and security needs—to an action, which corresponds to selecting particular $q$ and $N$ values.
To train the Q-learning agent, we initialize the Q-table with random values for all state-action pairs and update it iteratively using the Bellman equation \cite{bellman1957dynamic}. The agent selects new action $a'$ either through exploration (randomly selecting an action) or exploitation (choosing the action with the highest Q-value) based on an epsilon-greedy policy. The agent chooses a random action with exploration rate $\epsilon$, and the action that maximizes the Q-value with probability $1-\epsilon$. Specifically, the action is calculated as follows:

\begin{equation}
a' = 
\begin{cases} 
\text{random action} & \text{if } \text{random value} < \epsilon \\
\arg\max_{a'} Q(s', a') & \text{otherwise}
\end{cases}
\end{equation}
where $s'$ is the new state updated according to the chosen action $a'$.

To address \textbf{RQ3: Optimize}, we design the following reward function for RL in FL which balances utility gains and security: 

\begin{equation}
    R=\sum_{k=1}^{K}\sum_{i\in\mathcal{T}_k}\frac{\Delta Util_i}{|\mathcal{T}_k|L_k}-\alpha \mathbbm{1}(S_k<S_i)
\end{equation}
where $\Delta Util_i$ is the utility gain of client i for one round, $|\mathcal{T}_k|$ is the number of clients in cluster $\mathcal{T}_k$, $L_k$ is the round time of cluster $k$, $S_t$ and $S_i$ is the security level of cluster $k$ and security of client $i$, respectively. And, $\alpha$ is the security penalty. For a larger $\alpha$, the security loss increases, which result in lower rewards; conversely for smaller $\alpha$.

It is critical to note that rewards reflect the effectiveness of the chosen $q$ and $N$ parameters in balancing security and computational efficiency, addressing \textbf{RQ4: Effectiveness}. Higher rewards are given when selected parameters result in an optimal trade-off, maintaining strong encryption without excessive overhead, while lower rewards are assigned to suboptimal actions that compromise security or consume unnecessary resources. Over time, the agent shifts from exploration to exploitation, converging on the optimal action for each state by maximizing cumulative rewards and fine-tuning the HE parameters for each client sub-tier.

 Based on the current state, selected action, and corresponding rewards, Q-value is updated as:

\begin{equation}
Q(s,a) \leftarrow Q(s,a) + \gamma \left[ R + \mu \max_{a'} Q(s',a') - Q(s,a) \right]
\end{equation}
where $\gamma$ is learning rate, $\mu$ is the discount factor, the value of $Q(s,a)$ is obtained according to the Q-table. The detailed procedure is presented in Algorithm \ref{algo}.

\begin{figure}[!h]
    \centering
    \includegraphics[width=0.8\linewidth]{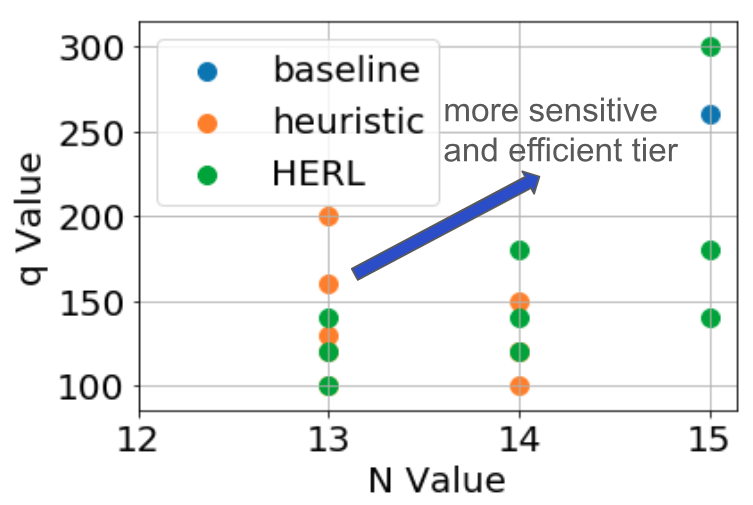}
    \caption{Selected parameter distribution of different approaches.}
    \label{fig:q-distribution}
\end{figure}
\begin{figure}
    \centering
    \subfigure[Convergence line of FL training w.r.t. different parameter plans.]{\includegraphics[width=.48\linewidth]{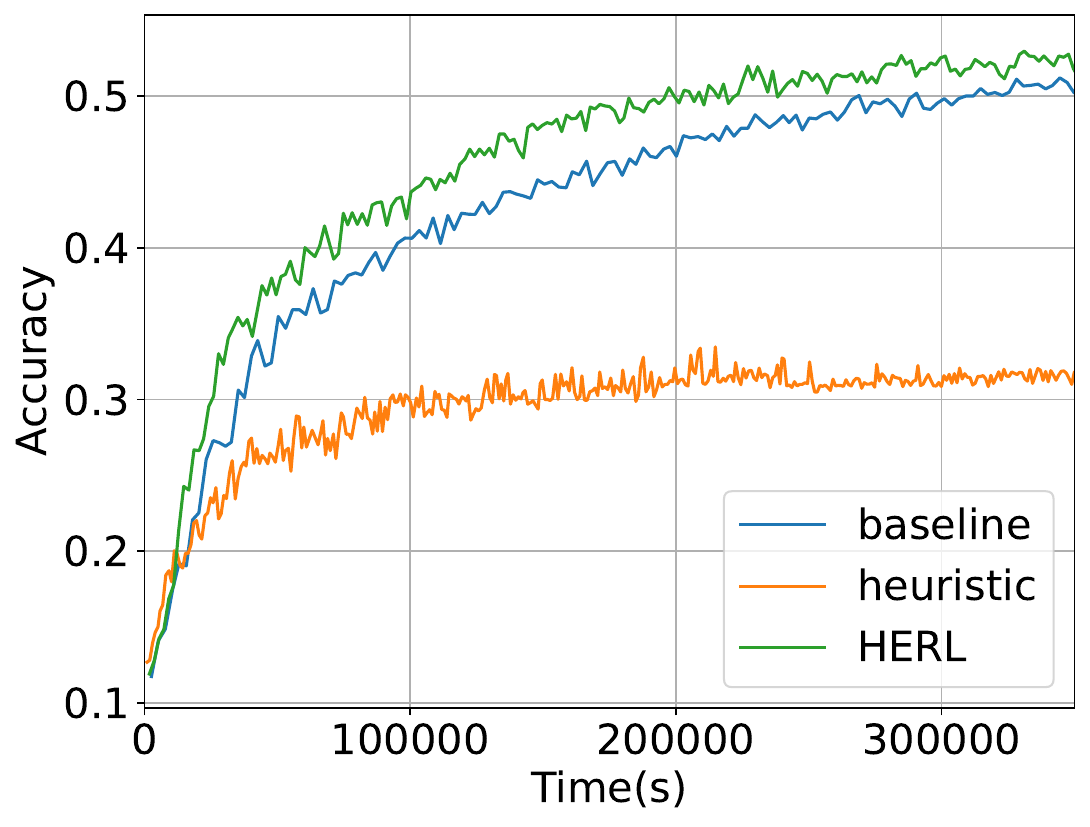}}\label{fig:q-converge}
    \subfigure[Breakdown performance.]{\includegraphics[width=.48\linewidth]{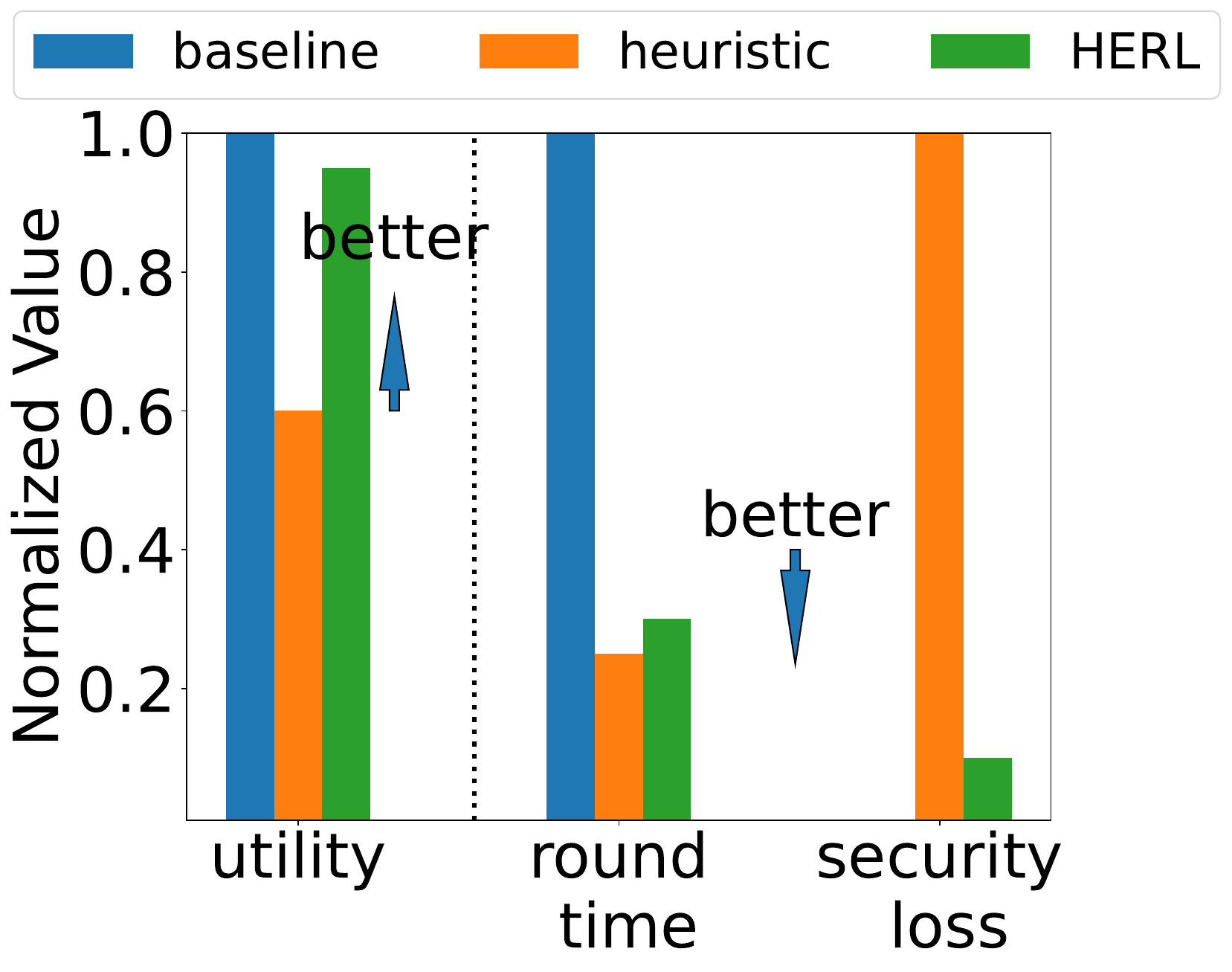}}\label{fig:breakdown}
    \caption{RL-agent performance (security loss is the number of clients not achieving the individual security requirement).}
    \label{fig:q-performance}
\end{figure}

The Q-learning approach is designed to be flexible and scalable, capable of adjusting to changes in client conditions efficiently, such as shifts in available resources or security demands. The reflection time of the RL-agent is less than 1s for each training round, which is efficient for large-scale FL systems. In the following sections, we use FL+HC to explore the performance of the proposed agent, comparing with the \textit{baseline} (uniform plan) and \textit{heuristic} (latency greedy) approach.

\begin{algorithm}
\caption{\scheme: RL-based Parameter Selection Algorithm for HE}
\normalsize
\begin{algorithmic}
\State \textbf{Input:} Client set $\mathcal{C}$, cluster number $K$, selection ration $r$, security parameter $\kappa$, security penalty $\alpha$, RL learning rate $\gamma$, discount factor $\mu$, exploration rate $\epsilon$, client $i$'s latency $L_i$, utility $U_i$, security requirement $S_i$.
\State \textbf{Output:} HE parameter plan for each tier $P = \{\mathcal{T}_k, q^k, N^k\}_{k=1}^K$.

\State \textit{/* Initialize global parameters. */}
\State $\mathcal{E} \gets \emptyset$ \Comment{Initialize clients and profiles}
\State $(pk_k, sk_k) \gets$ \textbf{KGen}($\kappa, \mathcal{T}_k$) \Comment{Generate key pairs for each tier}
\State Initialize Q-table $Q(s, a)$ with random values for each state-action pair.

\vspace{5pt}
\For{each round $j = 1, 2, \ldots$}

    \State Sample a set of participants $P^j = \{i | i \in \mathcal{C}\}$
    \State Profiling the information of all participants $\mathcal{E}^j=\{e_i=(U_i,L_i,S_i):i\in P^j\}$ \Comment{Participants of each round}

    \State $\{\mathcal{T}_k, k \in [K]\} \gets \Call{Tiering}{\mathcal{E}^j, P^j}$\Comment{Tiering the unexplored clients to the nearest tier}

    \State $P\leftarrow$\Call{ParamSelection}{$\{\mathcal{T}_k, k \in [K]\}$, $Q(s, a)$}
    \State Aggregation and global model broadcast.

\EndFor

\Function{ParamSelection}{$\{\mathcal{T}_k, k \in [K]\}$, $Q(S, A)$}
    \ForAll{$k = 1 : K$}
        \State /* \textit{Use Q-learning to select optimal $q^k$ and $N^k$ values:} */

        \State Observe the current state $S = (\{e_i,i\in\mathcal{T}_k\})$

        \If{random value $ < \epsilon$}
            \State Select action $a$ randomly \Comment{Exploration}
        \Else
            \State Select action $a = \arg\max_{A'} Q(s, a')$ \Comment{Exploitation, choose ($q^k, N^k$) pair with highest Q-value}
        \EndIf

        \State Execute action $a$ to set $q^k, N^k$

        \State Perform training and evaluate performance

        \State Observe new state $s'$ and reward $R$ for tier $k$

        \State Update Q-table: $Q(s, a) = Q(s, a) + \gamma \left[R + \mu \max_{a'} Q(s', a') - Q(s, a)\right]$

        \vspace{5pt}
        \State $P$.update$(\mathcal{T}_k, q^k, N^k)$
    \EndFor

    \State \Return $P$

\EndFunction

\end{algorithmic}

\label{algo}
\end{algorithm}

\noindent\textbf{Analysis of HERL performance.} We deliver the following case study to analyze how RL works and performs to address the point \textbf{RQ5: Trade-off}.
We set $K=9$ tiers and $2$ criteria (latency and security). After tiering, the RL-agent will iteratively respond in each training round. The final parameter selection of different approaches is presented in Figure \ref{fig:q-distribution}. We notice a trend of small parameter preference for the \textit{heuristic} approach and uniform large parameter plan for \textit{baseline}. However, \scheme could scale well according to the characters of tiers. Figure \ref{fig:q-performance} (a) shows the performance of convergence time and utility, where \scheme converges the fastest with a similar utility as \textit{baseline}. The detailed breakdown across utility, round time, and security is shown in Figure \ref{fig:q-performance} (b). The \textit{baseline} sacrifices latency to guarantee utility and security; the \textit{heuristic} approach instead sacrifices the utility and security to guarantee lower latency; however, \scheme makes a better trade-off decision between efficiency and security based on the RL rewards. More detailed evaluations are presented in the following section.

\section{Experiment}
In this section, we evaluate the performance of \scheme through comprehensive empirical experiments.

\subsection{Setup}
We begin by introducing the environment and details of our simulation framework.

\textbf{Environment. }To demonstrate the effectiveness of HERL, we performed comprehensive cross-device FL experiments utilizing NVIDIA RTX 4090 GPUs. Our simulations include extensive cross-device scenarios involving 1,000 clients for the FL tasks. In each training round, 100 clients are randomly chosen to participate in the FL training tasks, maintaining a consistent 10\% participation rate. 
Additionally, we used actual resource traces to assign computational capabilities to each client, thereby creating a realistic simulation of heterogeneous environments. Our experimental design is built upon the open-source PFL framework \cite{chen2022pfl}, which follows a conventional FL setup. This approach allows us to accurately represent the differences in device capabilities and data distributions.

\textbf{Datasets and Models.} We assess the performance of FL in cross-device environments through the following widely-used applications on benchmark datasets, Image Classification using a ResNet-18 model on CIFAR10 and CIFAR-100 \cite{CIFAR10}.

\textbf{Data Partitioning.} In our experiments, we simulate a realistic FL environment by distributing the data non-IID (non-Independent and Identically Distributed) among the clients. To achieve this, we utilize a Dirichlet distribution to split the dataset. This method allows us to mimic the inherent data variability found in real-world scenarios, where different clients may have varying distributions of data. The Dirichlet distribution enables the creation of skewed data partitions, where certain clients may have a concentration of specific classes, while others might have a more balanced distribution. This approach not only reflects the diversity of data across different devices but also presents a more challenging and representative benchmark for evaluating the performance and robustness of FL models in heterogeneous environments.

\textbf{Baseline.} 
We consider four popular clustering approaches in FL as baseline, and the details are as follows:
\begin{itemize}
    \item FL+HC \cite{briggs2020federated}: Clients are clustered following the methodology described in Section \ref{sec:cluster}, where hierarchical clustering is applied.
    \item TiFL \cite{chai2020TiFL}: Clients are tiered solely based on their past round completion times, improving the efficiency of client grouping.
    \item IFCA \cite{ifca}: Clients are randomly divided into $K$ equal groups to ensure a balanced distribution of resources.
    \item Auxo \cite{liu2023auxo}: Clients are clustered based on their utility distribution, with utility gain being the primary metric for defining this distribution.

\end{itemize}

\textbf{Implementation details.}
We build \scheme on top of the above baselines. We use the real-world traces \cite{refl, traces} (including computing and communication) to simulate the capacity of clients. We randomly initialize a security requirement for each client.  We utilize the default parameter settings outlined in the FL benchmark. For our experiments, the minibatch size is set to 10. The initial learning rate is configured to 5e-3 for the learning models. We apply the linear scaling rule to adjust the learning rate as the training progresses. We set the total tier number to be $K=9$. For Q-learning, we use learning rate $\gamma=0.1$, discount factor $\mu=0.9$, and exploration rate $\epsilon=0.1$.

\subsection{Improvement over Clustered FL}
In this section, we compare the performance of \scheme with the baseline methods to explore how much improvement \scheme can demonstrate.

\begin{table}[!ht]
    \centering
    \caption{Overall comparison among cluster FL approaches with HERL and Baseline.}
    \begin{tabular}{c|c|cc|cc}
    \hline
        \multirow{2}{*}{Dataset} & \multirow{2}{*}{Clustering}&\multicolumn{2}{|c|}{Final accuracy} & \multicolumn{2}{|c}{Converge time(s)}\\\cline{3-6}
        &&Baseline&HERL&Baseline&HERL\\\hline
         \multirow{4}{*}{CIFAR10}& FL+HC&0.47&0.52&289473&237584\\
         &TiFL&0.46&0.53&337483&258473\\
         &IFCA&0.46&0.48&317467&238746\\
         &Auxo&0.48&0.49&258373&201763\\ \hline
         \multirow{4}{*}{CIFAR100}&FL+HC&0.23&0.25&248617&192018\\
         &TiFL&0.27&0.26&266067&214996\\
         &IFCA&0.23&0.27&236843&218982\\
         &Auxo&0.23&0.25&267319&219260\\
         \hline
    \end{tabular}
    
    \label{tab:improvement}
\end{table}

\noindent\textbf{Overall performance.}
We summarize the performance in Table \ref{tab:improvement}. For \scheme 's performance on utility, it achieves improvements in final accuracy compared to baselines in most cases. On dataset CIFAR100 with IFCA clustering, it enhances the final accuracy up to 17.3\%. For the convergence time, the \scheme significantly reduces the convergence time for all scenarios by at least 15\%. On dataset CIFAR10 with TiFL clustering, the convergence time is reduced up to 24\%.

\begin{figure*}[!ht]
    \centering
    \subfigure[CIFAR10]{\includegraphics[width=0.4\linewidth]{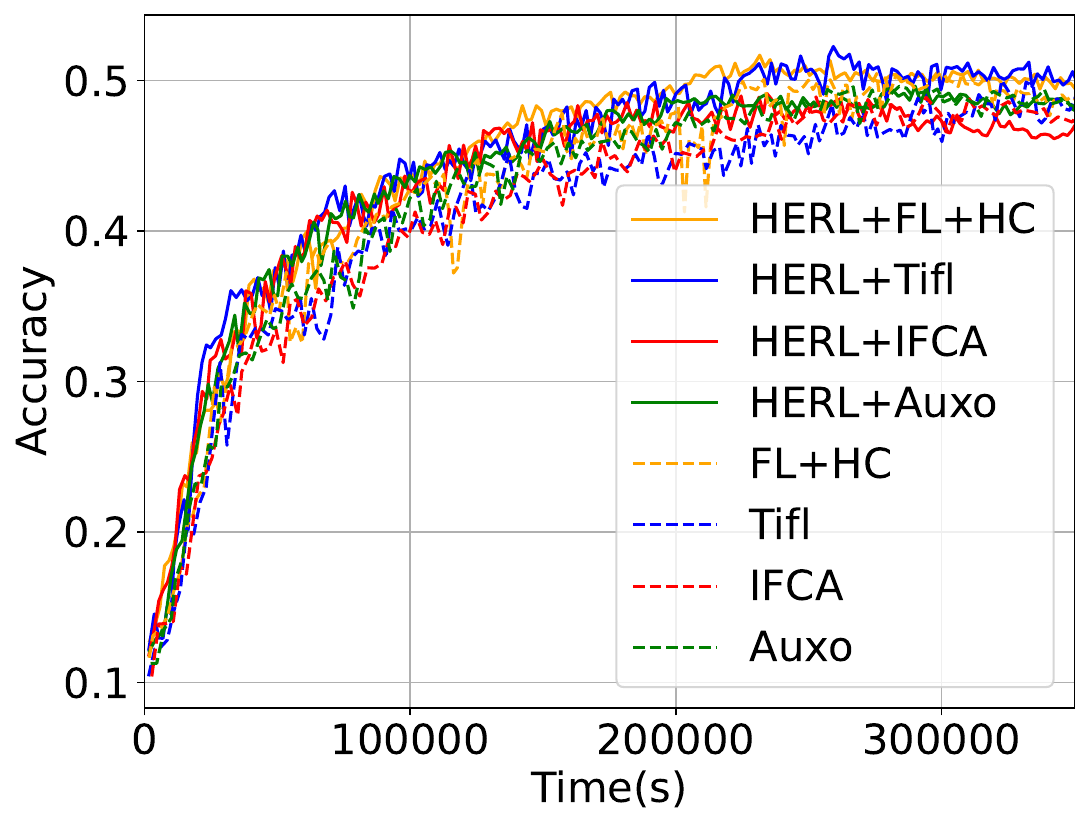}}
    \subfigure[CIFAR100]{\includegraphics[width=.4\linewidth]{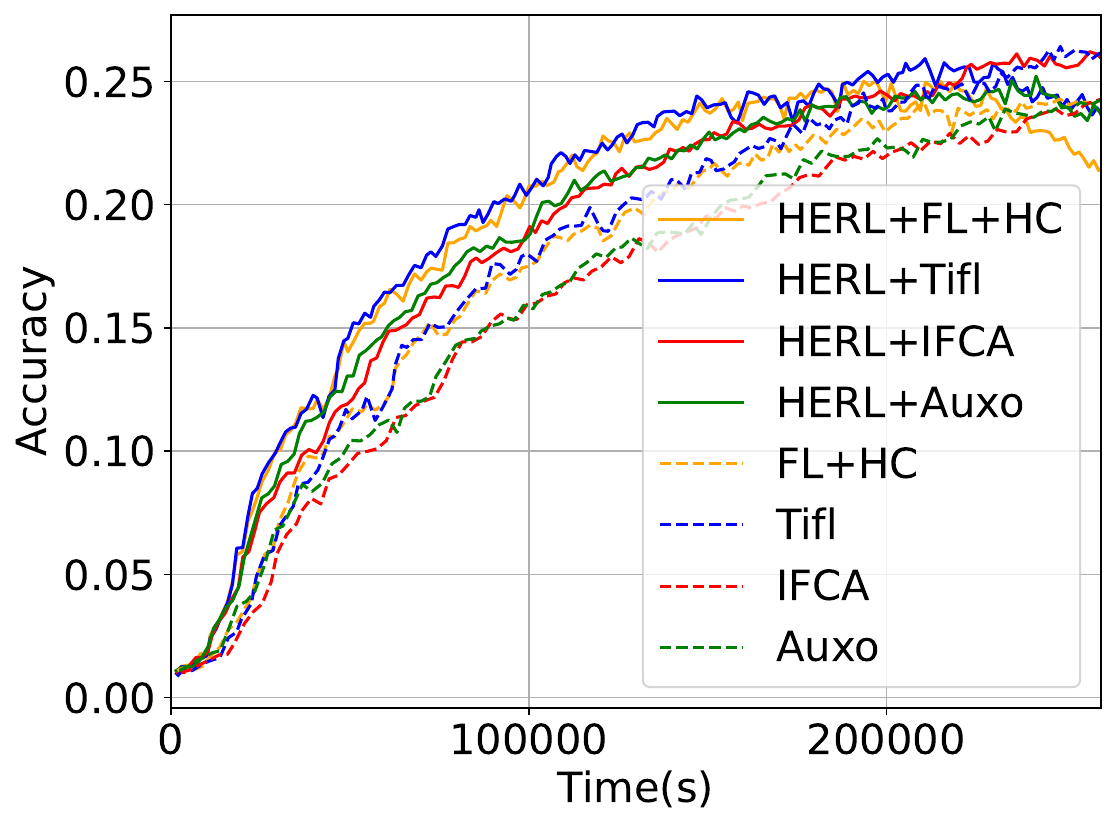}}
    \caption{Comparison on convergence line of HERL and \textit{baseline} approaches.}
    \label{fig:improvement}
\end{figure*}

\noindent\textbf{Efficiency.}
As shown in Figure \ref{fig:improvement}, we can observe the efficiency of the proposed \scheme in comparison to baseline approaches across both CIFAR-10 and CIFAR-100 datasets. In general, \scheme (solid line) consistently outperforms other baselines (dash line), showing faster convergence and achieving higher accuracy over time. Within the same time period, \scheme achieves up to 30\% more utility gain (or convergence efficiency) on CIFAR-100. The faster convergence of \scheme highlights the benefit of incorporating RL-based optimization for better adaptation to client heterogeneity, resulting in enhanced efficiency.

\begin{figure}[!ht]
    \centering
    \includegraphics[width=0.8\linewidth]{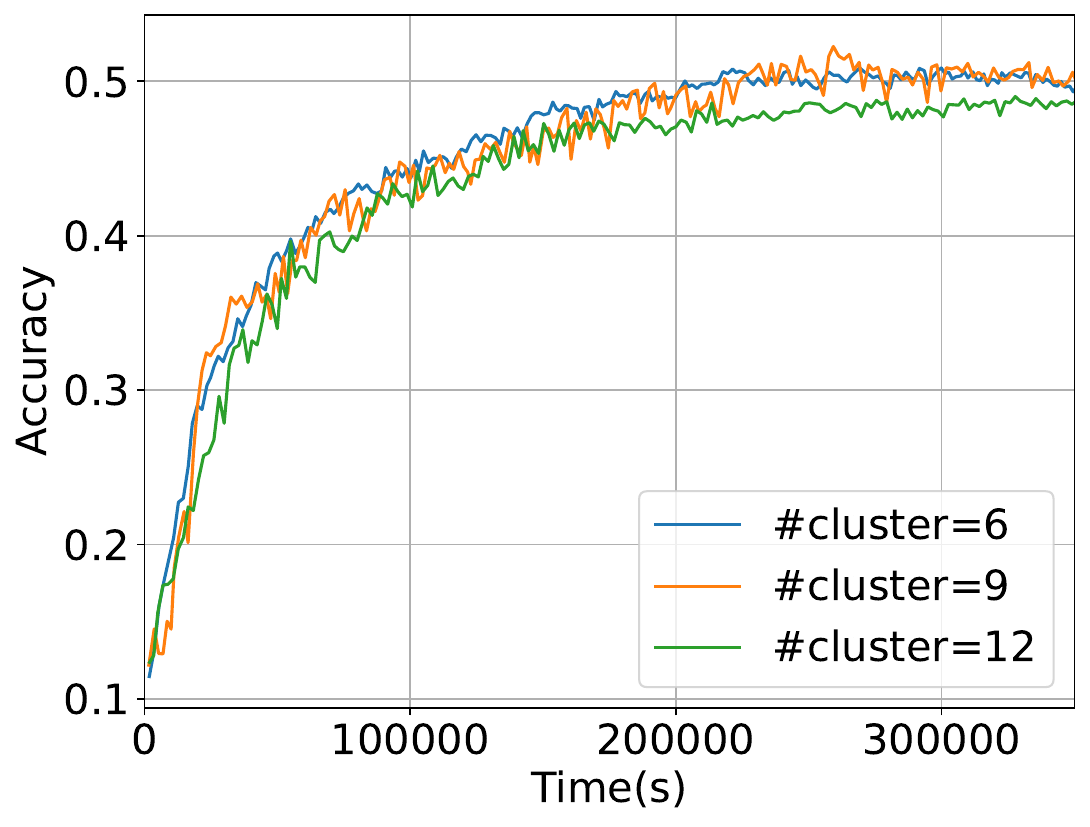}
    \caption{Influence of the number of clusters.}
    \label{fig:num_cluster}
\end{figure}

\begin{figure}[!ht]
    \centering
    \includegraphics[width=0.8\linewidth]{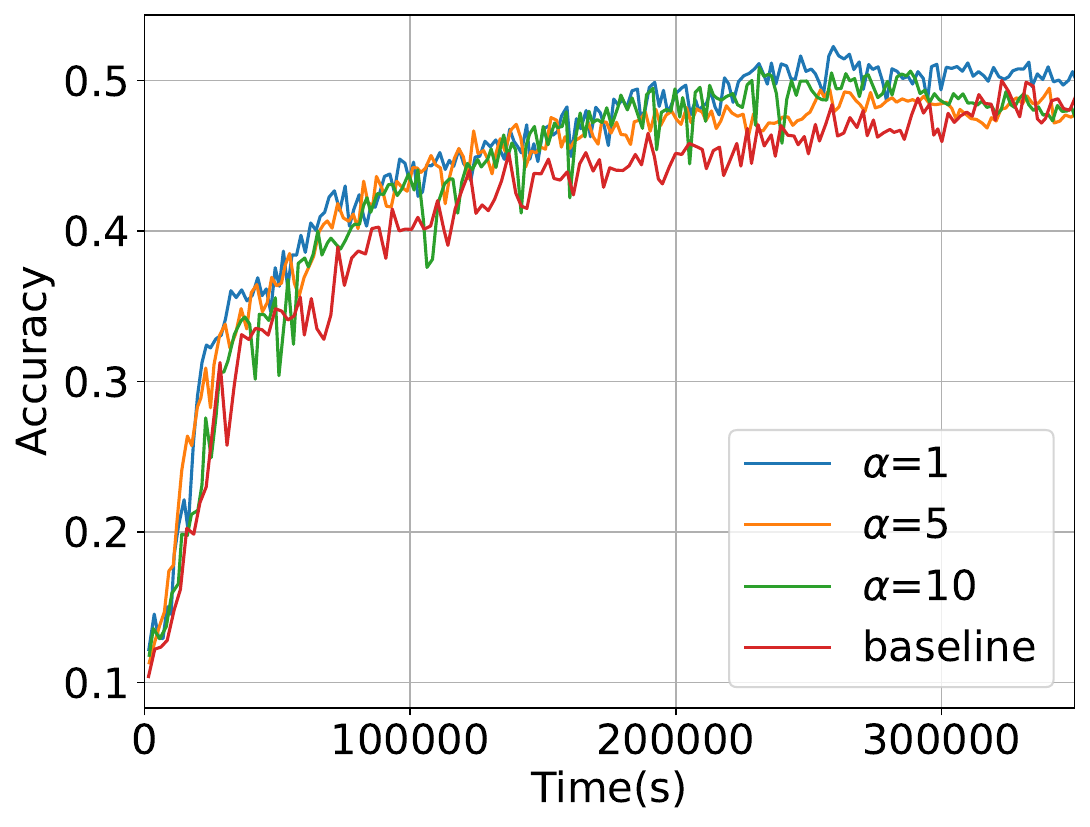}
    \caption{Influence of the number of clusters (\textit{baseline}: use the uniform recommended parameter plan).}
    \label{fig:security}
\end{figure}
\subsection{Sensitivity Analysis}
In this part, the sensitivity of hyper-parameters will be explored. We use the FL+HC as a base case to explore the relevant relationships.

\noindent\textbf{Impact of the number of clusters.} Figure \ref{fig:num_cluster} demonstrates the influence of varying the number of clusters on model performance using FL+HC clustering. The results show that increasing the number of clusters from 6 to 9 leads to improved accuracy and faster convergence. This suggests that a finer granularity in clustering helps the model better capture the heterogeneity of clients, allowing for more efficient aggregation of model updates. However, when the number of clusters is increased further to 12, a slight decline in performance is observed. This decline is likely due to increased overhead from managing more clusters, which may introduce communication and computational inefficiencies. These results highlight the existence of an optimal number of clusters, balancing the trade-off between fine-grained clustering and its additional overhead.
% it introduces.

\noindent\textbf{Impact of Security Penalty $\alpha$.} In Figure \ref{fig:security}, the effect of varying the security penalty factor ($\alpha$) on model accuracy is evaluated. A lower security penalty ($\alpha=1$) results in faster convergence and higher accuracy, as the reduced security overhead allows clients to allocate more resources to training, thereby improving performance. As the security penalty increases to $\alpha=5$ and $\alpha=10$, a gradual decrease in accuracy and convergence speed is observed. This is because higher security penalties impose more strict encryption and computational requirements, increasing resource consumption and slowing down the learning process. We could see the lines of \scheme gradually approach \textit{baseline} as the security penalty increases, since \textit{baseline} is set with a higher security level. This analysis shows the trade-off between security and convergence efficiency enabled by penalty $\alpha$.

\section{Conclusion}

In this paper, we proposed HERL, a novel RL-based approach for optimizing HE parameters in FL systems. HERL addresses the inherent challenges of client heterogeneity in FL by profiling and tiering clients according to specific needs. By employing Q-learning, HERL dynamically adjusts HE parameters, allowing for a tailored trade-off between computational efficiency, security, and utility.

Our experimental results demonstrate the effectiveness of HERL, showing significant improvements in both accuracy and convergence time across different clustered FL frameworks. Specifically, HERL outperformed baseline approaches, reducing convergence time by up to 24\% and improving accuracy by up to 17\%. These results demonstrate the potential of adaptive HE parameter selection in improving FL performance while maintaining data privacy and security. Future work can explore further optimization techniques and extend HERL to more complex and decentralized FL environments. This could further enhance the scalability and security of FL systems, addressing real-world challenges and limitations in FL.

% \newpage
\bibliographystyle{IEEEtran}
\bibliography{refs}

\end{document}